\documentclass[aps,prx,reprint,showpacs,superscriptaddress]{revtex4-1}
\usepackage[english]{babel}
\usepackage{amsmath,amssymb,bbm,graphicx,color,comment,txfonts}
\usepackage[bookmarks=true,colorlinks,citecolor=blue,urlcolor=blue]{hyperref}
\usepackage{dsfont}
\usepackage[normalem]{ulem}
\usepackage{physics}
\begin{document}

\date{\today}

\title{Revealing measurement-induced phase transitions by pre-selection}
\author{M. Buchhold}
\affiliation{Institut f\"ur Theoretische Physik, Universit\"at zu K\"oln, D-50937 Cologne, Germany}
\author{T. M\"uller}
\affiliation{Institut f\"ur Theoretische Physik, Universit\"at zu K\"oln, D-50937 Cologne, Germany}
\author{S. Diehl}
\affiliation{Institut f\"ur Theoretische Physik, Universit\"at zu K\"oln, D-50937 Cologne, Germany}

\begin{abstract}
Pushing forward the understanding of general non-unitary dynamics in controlled quantum platforms has been fueled by the recent discovery of measurement-induced phases and phase transitions. So far, these transitions remained largely elusive, since they are masked in standard quantum mechanical observables due to the randomness of measurement outcomes. Here, we establish a general scheme -- pre-selection -- to make them observable: The outcome randomness is broken explicitly by steering the system towards a representative state, which corresponds to one out of exponentially many possible measurement outcomes. Remarkably, this steering can be chosen so gently that the basic properties of the underlying measurement-induced transition, such as entanglement structure and critical exponents, are not modified. Pre-selection introduces a unique dark or absorbing state with macroscopic order, replacing the maximally mixed stationary state of the unconditioned measurement trajectory ensemble. This creates a link of measurement-induced phase transitions to new forms of quantum absorbing state transitions, which can be detected by standard means via a local order parameter. This insight further enables a quantum simulation strategy, determining the underlying universality class in state-of-the-art quantum platforms without measurement readout.
\end{abstract}

\maketitle
The dawn of noisy intermediate scale quantum platforms calls for a broader understanding of quantum dynamics -- it opens a realm beyond the paradigm of deterministic, unitary evolution generated by a Hamiltonian. Non-unitary resources enter as imperfections like decoherence and dissipation, but also through unprecedented local control and manipulation of quantum degrees of freedom, for example by measurements.

These platforms thus form instances of functional, driven open quantum matter. This fuels the quest for novel types of many-body behavior, without counterpart in Hamiltonian systems in thermal equilibrium. A case in point are the recently identified measurement-induced phase transitions (MIPT)~\cite{Fisher2018,Li2019b,Skinner2019}, which result from the competition of unitary dynamics and measurements~\cite{Jian2020,fan2020selforganized,alberton2021enttrans,turkeshi2021measurementinduced}. MIPTs are subtle, in that they come without an extensive order parameter witnessing them~\cite{gullans2019,Gullans2020,choi2020prl}. Instead, they are heralded by the amount of entanglement produced on each run~\cite{Schomerus2019,ippoliti2020}: Entanglement growth effected by a generic Hamiltonian is counteracted by disentangling local monitoring processes. This enables a sharp transition with universal entanglement scaling behavior in the thermodynamic limit~\cite{Zabalo2020,nahum2021prxq}.

Local measurements make the experimenter acquire knowledge about the quantum system. Ironically, despite this fact, MIPTs remain hard to detect: Constructing quantum mechanical observables, which could serve as their fingerprint, requires to conduct experiments with the same state
produced in every run. However, the degeneracy of the measurement outcomes (e.g. spin-up or -down in a local qubit measurement) pushes the ensemble of measurement trajectories into a maximally mixed state, erasing all information on individually prepared wave functions. Generalized, state dependent observables such as R\'enyi entropies~\cite{Elben,Hsieh2021a} are subject to the same reproducibility bottleneck. In principle, the information gained from the measurements could be used to remedy the situation: post-selecting on trajectories with the same measurement outcomes generates an ensemble of identical pure states from which observables can be extracted. This problem is, however, exponentially costly~\cite{Li2019b}, unless special conditions are met~\cite{Noel2022}. 

In this work, we show how MIPTs can be made observable, via replacing post-selection by \textit{pre-selection}. The core idea is to break the measurement degeneracy by replacing the random outcome state by a unique  representative wave function, which coincides with one of the exponentially many possible measurement outcomes. When the representative state features an extensive order parameter, the entanglement transition coincides with an ordering transition on the level of the density matrix, and thus becomes observable by standard means.  This is achieved by equipping the dynamics with a directionality in Hilbert space. This steering can be performed such that the entanglement scaling and the stationary critical properties of the representative state quantitatively represent the entire ensemble, in this sense analogous to post-selection.
The phenomenology established below reveals that this scheme grants access to an intriguing new class of \emph{quantum absorbing state phase transitions}~\cite{Marcuzzi2016,Lesanovsky2019,Carollo2022,Carollo2019}. 

The concept of pre-selection is applicable to any MIPT. We exemplify its generality in fermion and spin models, with and without conservation laws, and for continuous and projective measurements. It comes in various incarnations, with and even without the need to read out the measurements. Yet all approaches share in common to give quantitative access to the universal properties of the underlying MIPT. 

\begin{figure*}[t]
  \includegraphics[width=1\linewidth]{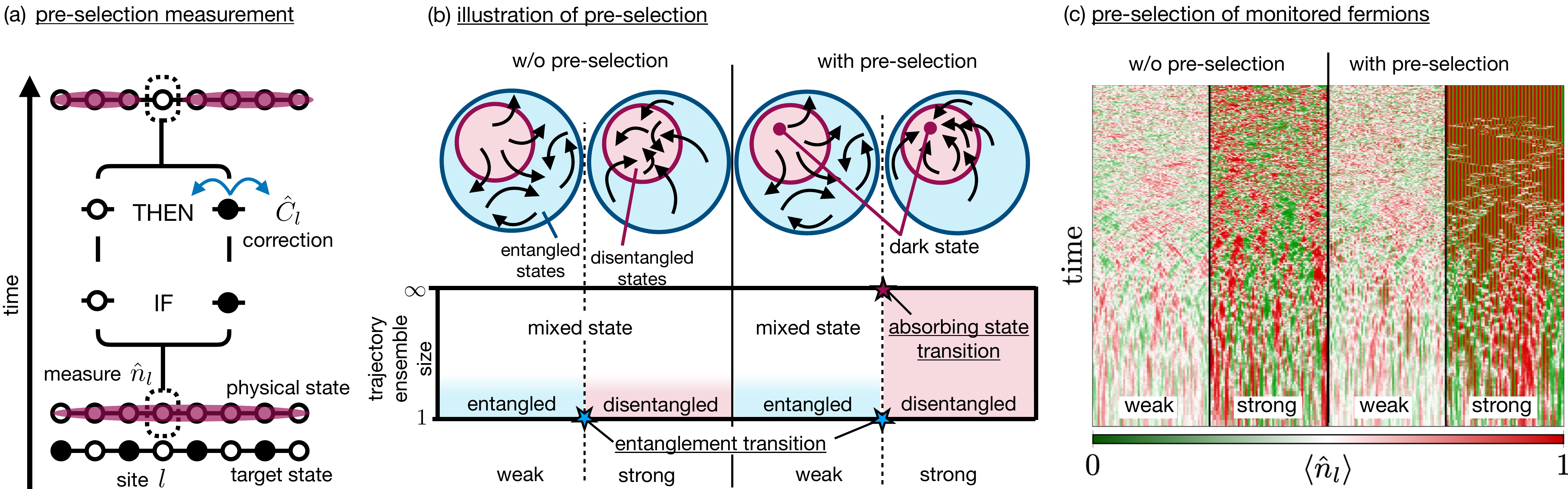}
  \caption{Pre-selection concept: (a) Illustration in a fermion hopping model with local monitoring. The outcome of measurements of $\hat n_l$ are recorded, and action is taken conditioned on its compatibility with a fixed target state by an IF-THEN clause discussed in the main text. Here a N\'eel state is targeted. If the occupation matches the one of this state, no feedback is applied. If the correlation is incompatible with the Néel state, a unitary hopping process is imposed.
  (b) General mechanism. Wave functions undergoing Hamiltonian dynamics subject to local measurements display two different dynamical phases: for weak (rare) measurements, the Hamiltonian scrambles information and each initial wave function evolves into a (sub-) extensively entangled state. For strong (frequent) measurements, each wave function enters a quantum Zeno regime and obeys area law entanglement. Despite their different entanglement structure, the wave functions in both phases are subject to randomness in space and time, and their ensemble always approaches a maximally mixed state. In the presence of pre-selection, a unique dark state is introduced. The presence of the dark state does not affect the dynamics in the weak measurement phase, but due to feedback, it acts as an attractor of disentangled wave functions in the quantum Zeno regime. Each disentangled wave function evolves towards the dark state, making the ensemble a pure, disentangled state. (c) Unconditioned vs. pre-selected evolution for monitored free fermions. Pre-selection successfully distinguishes the weak and the strong measurement phase: for weak measurements each fermion wave function obeys a logarithmic entanglement growth and a random density distribution. For strong measurements, each state evolves into a N\'eel state with alternating particle number distribution. Without pre-selection, both weak and strong measurements yield a random state, which can only be characterized by its entanglement entropy.}
  \label{Fig1}
\end{figure*}

\textit{Pre-selection} -- The goal is to pull the MIPT to the observable level by modifying the dynamics such that it uncovers a phase transition in the density matrix $\hat \rho$, averaged over the measurement outcomes. In general, the averaged update of a state in a time step $\delta t$ can be written as a dynamical Kraus map  
\begin{eqnarray}\label{eq:kraus}
 \hat \rho_{t+ \delta t} = \hat{\mathcal X} [\hat \rho_t] \delta t,
\end{eqnarray}
with a time-local, completely positive, and trace preserving generator $\hat{\mathcal X}$. We denote the generator for the unmodified averaged unitary and measurement dynamics by $\hat{\mathcal X}^{(0)}$. Generically, the evolution has only one stable dynamical fixed point, namely, the maximally mixed state $\hat \rho \propto \hat{\mathds{1}}$, which obeys $\hat{\mathcal X}^{(0)} [\hat{\mathds{1}}] =0$. This is readily verified in the temporal continuum limit $\delta t \to 0$, where the generator takes Lindblad form, 
\begin{eqnarray}\label{eq:lind}
 \hat{\mathcal X} [\hat \rho]  = - i [\hat H , \hat \rho ] + \sum_l \hat L_l  \hat \rho \hat L_l^\dag -\tfrac{1}{2} \{ \hat L_l^\dag\hat L_l, \hat \rho\},
 \end{eqnarray}
with $l$ labelling lattice sites. We consider measurements of local operators $\hat L_l^{(0)}$ with eigenvalues $\{\lambda_1,\lambda_2\}$ spanning a two-dimensional Hilbert space. This yields the unmodified Hermitian Lindblad operators
\begin{align} 
\hat L_l^{(0)} = (\hat L_l^{(0)})^\dag=\lambda_1\hat P_l+\lambda_2(\hat{\mathds{1}}-\hat P_l), \text{ with }\hat P_l=|{\lambda_1}\rangle\langle{\lambda_1}|,
\end{align}
with $\hat P_l^2= \hat P_l$. 
Examples are measurements of the spin orientation $\hat Z_l \,\,(\lambda_{1,2} = \pm 1)$ with $\hat P_l=(\hat{\mathds{1}}- \hat Z_l)/2$ or fermion particle number $\hat n_l \,\,(\lambda_{1,2} = 0,1)$ with $ \hat P_l= \hat{\mathds{1}}-\hat n_l$.

In order to make the transition observable, we will introduce modifications of the generator  $\hat{\mathcal X}^{(0)} \to \hat{\mathcal X}$ ($\hat H^{(0)}, \hat L^{(0)}\rightarrow \hat H, \hat L$) such that the following three conditions are obeyed:\\
(c1) There exists a \textit{dark state} $\hat \rho_D=|D\rangle\langle D|$ of the full generator, i.e., $\hat{\mathcal X}[\hat \rho_D ] = 0$, which represents one possible measurement outcome.\\ 
(c2) The dark state is the unique stationary solution.
\\
(c3) All modifications preserve the symmetries of the original problem and are irrelevant in the renormalization group (RG) sense.
 
Condition (c1) pre-selects a representative state of the measurement outcomes, and it can be equipped with an extensive order parameter, which eases experimental detection in many-body systems. Without loss of generality, we associate the projectors $\hat P_l$ with the `right' measurement outcome, which is compatible with the dark state $\hat P_l\hat\rho_D=\hat\rho_D$. For local measurements, $\hat\rho_D$ is a product state. (c2) introduces a directionality in the evolution in Hilbert space: iterating the dynamical map  ultimately filters out the state $\hat \rho_D$. In particular, the maximally mixed state ceases to be a dynamical fixed point of the averaged evolution. (c3) ensures that the modified dynamics is consistent with the original transition, and its universality class remains unaltered.

\begin{figure*}[t]
  \includegraphics[width=1\linewidth]{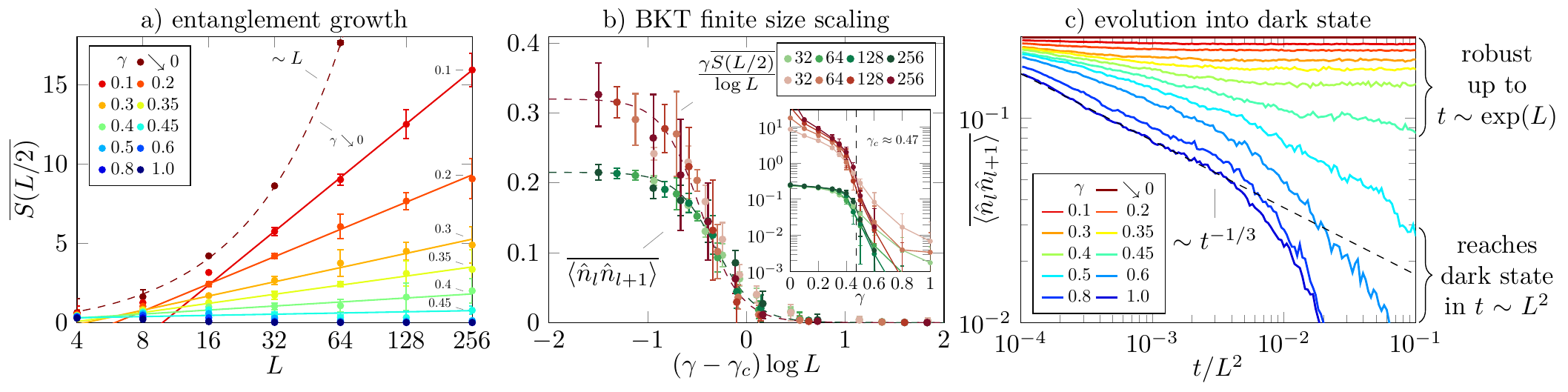}
  \caption{Absorbing state phase transition of monitored free fermions with pre-selection. (a) Growth of the half-chain entanglement entropy $S(L/2)$ with system size $L$. In the weak measurement phase ($\gamma\lesssim0.47$), the entanglement grows logarithmically $S(L/2)\sim \tfrac{c_\text{eff}}{3}\log L$, and it vanishes in the strong measurement phase ($c_\text{eff}=0$) since the entanglement entropy in the Néel state is exactly zero (data points for $\gamma>0.47$ overlap). In the limit of vanishing measurement strength, a volume-law growth of the entanglement is observed. (b) Finite size scaling of the entanglement entropy and the ensemble averaged density-density correlation function $\langle \hat{n}_l \hat{n}_{l+1}$, extracted after an evolution time of $t=0.1L^2$. The scaling collapse reveals that both the entanglement entropy and the ensemble order parameter indicate a phase transition at the same critical measurement strength $\gamma_c=0.47$. Here, the exact vanishing of $S(L/2)$ and $\langle\hat n_l\hat n_{l+1}\rangle$ for $\gamma>0.47$ reduces the uncertainty in determining $\gamma_c$ compared to the unmodified MIPT, which was reported at a comparable value of  $\gamma_c=0.31$~\cite{alberton2021enttrans}. The data  displays BKT essential scaling with an exponential divergence of the correlation length $\log\xi\sim|\gamma-\gamma_c|^{-1}$. This demonstrates the connection of the absorbing state phase transition in the ensemble average and the entanglement transition on the level of individual trajectories. At the critical point and in the thermodynamic limit, both the order parameter and the entanglement entropy undergo a discontinuous jump from $\langle \hat{n}_l \hat{n}_{l+1} \rangle \approx 0.21$ or $ c_\text{eff} \approx 2.0$ to zero. The inset shows the unrescaled data, with dashed lines as guides to the eye. (c) The evolution of the order parameter $\langle \hat n_l \hat n_{l+1}\rangle$ displays a characteristic behavior in both phases: in the strong measurement phase, the dark state is reached after an algebraic time scale $t\sim L^2$ resulting from the diffusive redistribution of number-conserved particles under strong measurements. The initial relaxation follows a power law decay $\sim t^{-1/3}$ of the order parameter. In the weak measurement phase, the order parameter saturates on a plateau, which is robust up to exponentially long time scales $t\sim \exp(L)$ in the system size. This phenomenology in the dynamics gives a further parallel to absorbing state transitions, and to purification transitions~\cite{gullans2019}.}
  \label{Fig2}
\end{figure*}

These three conditions can be implemented in various distinct ways. Their common thread is to condition the evolution at a given time step on the measurement outcome of the previous one, in the sense of an `IF-THEN' clause: if a certain measurement outcome is obtained, then depending on the compatibility with the target state $\hat \rho_D$ the state update is selected. In particular, below we expose three principal strategies to implement this goal, which we dub `classical feedback',  `quantum feedback' and `quantum simulation'. 
 
The feedback strategies rely on reading out the measurement. In the classical version, the Hamiltonian $\hat H^{(0)}\to \hat H[\hat \rho]$ is chosen conditioned on the detected state $\hat \rho$ such that steering in Hilbert space is achieved, indeed akin to feedback \cite{Wiseman1993,Wiseman1994,
Ivanov2020}. The evolution then becomes non-linear in the state. The alternative quantum strategy builds on correcting `wrong' measurement results immediately~\cite{Gefen,Santos_2020}: it modifies $\hat L^{(0)}\to \hat L$ by sending $(\hat P, \hat{\mathds{1}}-\hat P) \to (\hat P, \hat C (\hat{\mathds{1}}-\hat P))$ with unitary correction $[\hat C, \hat P]\neq 0$. 
The quantum simulation approach refrains entirely from surfacing the measurement outcomes $\hat L^{(0)}$ to the experimenter's classical world: Instead, internal quantum action resulting from coupling to an environment is taken~\cite{Diehl2008,Verstraete2009,Weimer2010}, but here conditioned on the effect of the operator $\hat  L^{(0)}$. This relaxes the constraints -- measurements need not be read out and $\hat C$ need not be unitary -- without modifying (c1)-(c3) or the phase transition on the average ensemble.
In the quantum strategies, the evolution remains linear in the state, but the Lindblad operators $\hat L$ become non-Hermitian, implementing the desired directionality in Hilbert space. Thus they bear some similarities with the scenario of dark state cooling and associated phase transitions~\cite{Diehl2010,Marino2016}, with the key qualitative difference that these usually do not feature a dark state in the entire parameter space.

How can a phase transition then occur, if the dark state is a stationary solution globally in parameter space? An insight can be obtained from a parallel to absorbing state phase transitions, such as directed percolation~\cite{Hinrichsen_2000,RevModPhys.76.663}. There as well, exists a global absorbing state. However, for generic initial conditions, it is only reached below a certain threshold, while it will be missed by the dynamics (for times sub-exponential in the system size) above it~\cite{Hinrichsen_2000}. This defines absorbing and active phases, respectively, and there is a critical point separating them. The transitions here can be manifestly in the quantum regime, witnessed, e.g., in a $1+1$ dimensional example below, reminiscent of a zero temperature quantum phase transition.

\emph{MIPT for monitored free fermions.} -- We illustrate each pre-selection scheme for a MIPT of monitored free fermions~\cite{alberton2021enttrans}. The fermion Hamiltonian describes hopping on a half-filled chain of length $L$ (units are set by the Hamiltonian)
\begin{align}\label{eq:FeedbackHam}
    \hat H^{(0)}= \sum_{l}\hat h_l r_l,
\end{align}
with $\hat h_l=\hat c^\dagger_l \hat c_{l+1}$+H.c. for fermions $\hat c^\dagger_{l},\hat c_{l}$, and $r_l=1$ is the coupling parameter here. The Hamiltonian competes with measurements, either weak-continuous or projective, of non-commuting observables $[\hat L_l^{(0)},\hat h_l]\neq0$. Here, we measure the local densities $\hat L^{(0)}_l=\hat n_l=\hat c^\dagger_l \hat c_l$ with rate $\gamma$.

This competition drives a MIPT between two different entanglement phases~\cite{alberton2021enttrans,buchhold2021effective}: for small rates $\gamma$ (weak measurements), each wave function is scrambled and approaches a state with logarithmic entanglement growth, reminiscent of a conformally invariant, quantum critical state. For large $\gamma$ (strong measurements), a quantum Zeno regime emerges. It hosts area law entangled wave functions, close to instantaneous eigenstates of each $\hat L^{(0)}_l$. Both regimes are separated by a measurement-induced Berezhinskii-Kosterlitz-Thouless (BKT) phase transition~\cite{buchhold2021effective,bao2021symmetry}. 

\emph{Classical pre-selection.} -- In order to lift the measurement-degeneracy, pre-selection is implemented by making $\hat H^{(0)}\to \hat H[\hat\rho]$ state-dependent via \begin{align}
    r_{l}\rightarrow r_{l}[\hat\rho] \text{ but leaving} \ \hat L_l\equiv \hat L_l^{(0)}. 
\end{align} \label{eq:Feedbackroutine}
A suitable absorbing state with $[\hat H[\hat\rho_D],\hat \rho_D]=0$
is the Néel state $|D\rangle=|01010...\rangle$,
populating either the even or the odd sites of the chain, see Fig.~\ref{Fig1}(d). A convenient choice for the state-dependence of $r_l[\hat\rho]$ is
\begin{align}
    r_l[\hat\rho]= 2-(\lfloor\langle \hat n_{l-1}
    \rangle\rfloor-\lfloor\langle\hat n_{l}\rangle\rfloor)^2-(\lfloor\langle \hat n_{l+1}
    \rangle\rfloor-\lfloor\langle\hat n_{l+2}\rangle\rfloor)^2,\nonumber
\end{align} 
where $\lfloor...\rfloor$ indicates rounding to the nearest integer. This conditions the couplings $r_l$ on the instantaneous expectation values $\langle \hat L_l^{(0)}\rangle$ of the measured operators. In an experiment, these can be obtained by simulating the instantaneous state $\hat \rho$ (e.g., for Gaussian systems or stabilizer states, see Appendix~\ref{app:continuous_monitoring}) or be extracted from the measurement outcomes, e.g., by applying an appropriate noise filter~\cite{Young_2021}.

An intuitive understanding obtains in the limiting cases: (i) For weak measurements, the operators $\hat h_l$ scramble the observables $\hat n_l$. The state becomes translation invariant $\langle\hat n_l\rangle=1/2$ and $r_l[\hat\rho]\rightarrow1$  is effectively state-independent. (ii) For strong measurements, $\hat\rho$ enters a quantum Zeno regime, where $\langle \hat n_l\rangle\hat\rho\sim \hat n_l\hat\rho$. We can thus promote $r_{l}[\hat\rho]\rightarrow \hat r_{l}\equiv 2-(\hat n_{l-1}-\hat n_{l})^2-( \hat n_{l+1}-\hat n_{l+2})^2$ to an operator with $r_{l}[\hat\rho]\hat\rho=\hat r_{l}\hat\rho$. A directed evolution towards $\hat\rho_D$ is implemented when the product $\hat h_l \hat r_l$ is non-Hermitian, i.e., when $[\hat h_l, \hat r_l]\neq0$. Both limits, (i) and (ii), allow us to effectively eliminate the state dependence in $\hat H[\hat\rho]$, yielding two types of Hamiltonians $\hat H_{w,s}$: for weak measurements $\hat H_{w}\equiv\hat H^{(0)}$ and thus is unmodified by pre-selection. For strong measurements $\hat H_s\equiv \sum_{l}\hat h_l \hat r_l$ is non-Hermitian and implements an IF-THEN condition, steering towards $\hat\rho_D$.

Fig.~\ref{Fig2} demonstrates the MIPT with pre-selection for a continuous-measurement scheme (see Appendix~\ref{app:continuous_monitoring}): The entanglement scaling is inherited in the pre-selected case, but now accompanied by a phase transition for the order parameter $\sum_l\langle \hat n_l\hat n_{l+1}\rangle$. The critical behavior is governed by the essential scaling of a BKT transition. The  transition can be probed experimentally by  standard means, like a single-shot measurement of the density distribution.

\emph{Quantum pre-selection.} -- We modify the Hamiltonian in Eq.~\eqref{eq:FeedbackHam} such that the N\'eel state is again a dark state: 
\begin{align}
\label{eq:condFerm}
\hat H^{(0)}& \rightarrow \hat H=\sum_{l \text{ odd}}\hat n_l (\hat h_{l+1}+\hat h_{l-2})+\sum_{l \text{ even}} (1-\hat n_l) (\hat h_{l+1}+\hat h_{l-2}).\nonumber
\end{align}
The directionality is implemented by an IF-THEN condition acting right after a `wrong' (incompatible with target dark state) measurement result $\hat n_{l\,\text{odd}}\to 1$ has been obtained on odd sites. Wrong results can be corrected by applying a quasi-local unitary gate operation $\hat C_l$ with $[\hat C_l,\hat L^{(0)}]\neq 0$. Here we take $\hat C_l=\exp(-i\tfrac{\pi}{2}\hat h_l)$. This preserves the symmetries, and works both for projective measurements and quantum jump~\cite{Daley} evolution protocols (see Appendix~\ref{app:quantum_preselection}). The principle of the steering towards the dark state is this: If a `right' measurement result has been recorded, the projector $\hat P_l$ evolves the state towards $\hat\rho_D$. If a `wrong' measurement result has been recorded, the state is projected by $\hat Q_l=\hat{\mathds{1}}-\hat P_l$ away from $\hat\rho_D$. We then rotate it back onto the dark state by the unitary $\hat C_l$: 
\begin{align}
    \text{`right'}: \hat\rho\rightarrow \hat P_l\hat\rho\hat P_l / \langle \hat P_l\rangle, \ \text{`wrong'}: \hat\rho\rightarrow \hat C_l \hat Q_l \hat\rho\hat Q_l\hat C_l^\dagger/\langle \hat Q_l\rangle.\nonumber
\end{align}
The average ensemble of pre-selected wave functions then follows the Lindblad master equation (see Appendix~\ref{app:quantum_preselection})
\begin{equation}\label{eq:modifiedmaster}
    \partial_t\hat \rho=-i[\hat H,\hat \rho]-\frac{\gamma}{2}\sum_l\{[\hat n_l,[\hat n_l, \hat \rho]] +\{\hat L_l^\dagger\hat L_l,\hat \rho\}-2\hat L_l \hat \rho \hat L_l^\dagger\},
\end{equation}
with Lindblad operators $\hat L_l$ on odd/even sites $l$
\begin{align}
    \text{ odd: }\hat L_l=\hat C_l\hat L^{(0)}_l=\hat n_l\hat n_{l+1}+i\hat c^\dagger_{l+1}\hat c_l, \ 
    \text{ even: }\hat L_l=\hat L^{(0)}_l=\hat n_l.\nonumber
\end{align}
The master equation~\eqref{eq:modifiedmaster} turns the fermion MIPT in individual wave functions into an absorbing state transition for the ensemble: The repeated application of both $\hat H$ and $\hat L_{l\text{ odd}}$ pushes any initial state into the Néel state $\hat\rho_D$, which is the unique dark state. Active and inactive phases are distinguished by the characteristic time scale at which $\hat\rho_D$ is reached. For strong measurements ($\gamma\gg1$), the state $\hat\rho$ is nearly diagonal in the fermion number basis and the modified Lindblad operators cause a diffusive motion towards the Néel state (reached on time scales $t\sim L^2$ with system size $L$). The absorbing property of the Néel state in this regime is confirmed by a mean-field analysis of the Lindbladian for a homogeneous system of infinite size (see Appendix~\ref{app:meanfield}). It predicts that the Néel state is the only stationary state for $\gamma\ge2$. For weak measurements $\gamma\ll1$, however, the enhanced superposition of particles leads to a drastic enlargement of the accessible Hilbert space, and a slowing down of the relaxation towards the Néel state. It can only be reached on time scales growing with the size of the Hilbert space $t\sim\exp(L)$. In the mean-field equations, this is revealed by the emergence of a second stationary solution: an active state with $\langle \hat n_{l\text{ odd}}\rangle=\langle  1-\hat n_{l\text{ even}}\rangle=\tfrac{4-\gamma^2}{8}$ for $\gamma<2$. While the Néel state remains stationary, any other initial configuration approaches the active state, whose lifetime diverges in the $L\to\infty$ limit. This confirms the phenomenology of an absorbing state phase transition. In order to examine the long-wavelength dynamics beyond mean-field, we work in the spatial continuum limit, which for number-conserving fermion dynamics amounts to the bosonization framework~\cite{buchhold2021effective}. In bosonization, the irrelevance of the modifications leads to $\hat H \approx \hat H^{(0)} $ and $ \hat L_l\approx \hat L_l^{(0)}$ up to (globally crucial) gradient terms. Quantitatively, the dynamics is described by a non-Hermitian Sine-Gordon theory, which predicts the absorbing state transition to be BKT. This complements analytically our numerical results for the classical scheme (Fig.~\ref{Fig2}). In particular, the term describing the noise level in the effective theory is RG irrelevant, reflecting the existence of a dark state and preserving the scaling behavior characteristic of a $1+1$ dimensional quantum BKT scenario (see Appendix~\ref{app:fieldtheory}).

\begin{figure*}[t]
  \includegraphics[width=1\linewidth]{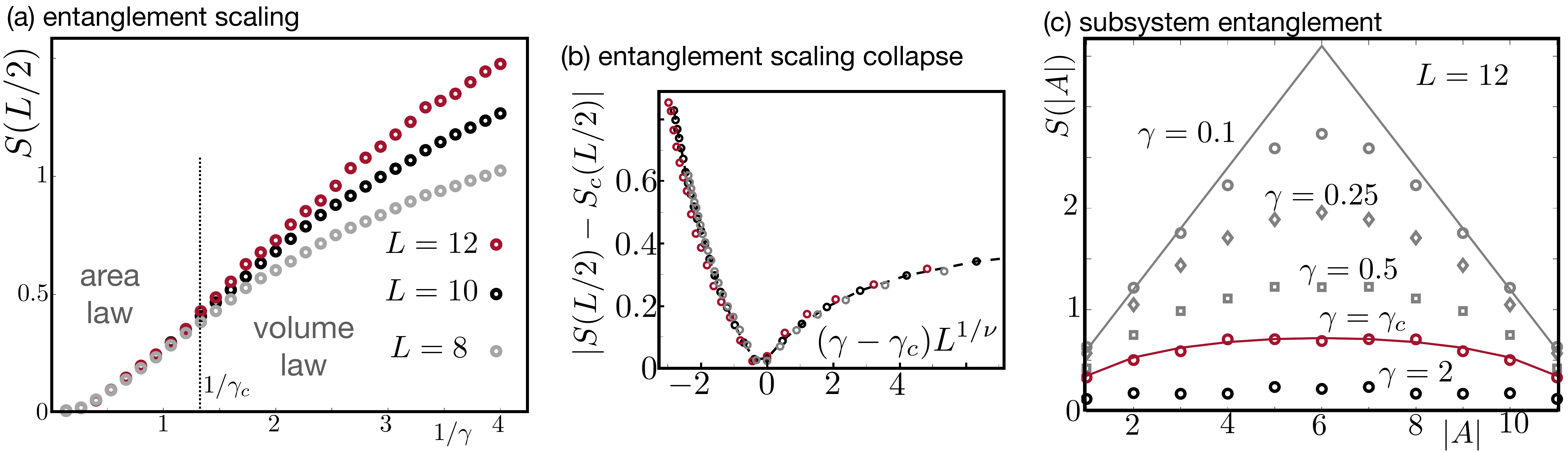}
  \caption{Measurement-induced phase transition in a Rydberg array. We numerically simulate the time evolution of an array of $L$ qubits, generated by the Rydberg Hamiltonian in Eq.~\eqref{eq:RydbergHam} and projective measurements of the orientation $\hat Z_l$ of each qubit. Changing the rate $\gamma$ at which measurements are performed, the system undergoes an entanglement phase transition in the stationary state. It is witnessed in the von Neumann entanglement entropy $S(|A|)=-\text{tr}(\hat\rho_A \ln \hat\rho_A)$, where $\hat\rho_A=\text{tr}_{\bar A}|\psi\rangle\langle\psi|$ is the density matrix of a contiguous subsystem $A$ after tracing out its complement $\bar A$. (a) For small measurement rates (weak measurements), the entanglement entropy for different system sizes obeys a volume law $S(L/2)\sim L$. For large measurement rates (strong measurements) instead, it collapses onto an area law $S(L/2)\sim \text{const.}$ Both regimes are separated by a critical point at $\gamma_c\approx0.77$. (b) Finite size scaling of the entanglement entropy is consistent with a correlation length critical exponent $\nu=\tfrac43$ for bond percolation, similar to what has been found in Ref.~\cite{Zabalo2020} for a related MIPT in Haar- and Clifford random circuits. (c) The entanglement entropy for a total system size $L=12$ and different subsystem sizes $|A|$ reveals the MIPT from volume- to area law. For weak measurements at $\gamma=0.1$, it displays a volume law with a nearly perfect linear growth $S(|A|)\approx0.6\min\{|A|,L-|A|\}$,  (grey solid line). At the critical value $\gamma=\gamma_c$, it is well described by a logarithmic function $S(|A|)=0.56\ln(\sin(\pi|A|/L))+s_0$ as in conformal field theory. For strong measurements ($\gamma=2$) it approaches a constant which is independent of $|A|$.}
  \label{Fig3}
\end{figure*}

\textit{Quantum simulation.} -- An even simpler practical route to extract the quantitative properties of the MITP is when an engineered environment~\cite{Diehl2008,Verstraete2009,Barreiro_2011} implements the IF-THEN condition autonomously without the need of measurement readout. Indeed, the master equation~\eqref{eq:modifiedmaster}  reveals a general equivalence between MIPTs with pre-selection and absorbing state phase transitions in open systems: whenever the Lindblad operators $\hat L_l$ can be written as a projector $\hat{\mathds{1}}-\hat P_l$ (IF) followed by an operator imposing directionality $\hat C_l$ with $[\hat C_l ,\hat L_l]\neq 0$ (THEN), the corresponding absorbing phase transition has a MIPT counterpart in monitored systems. When the operator $\hat C_l$ is imposed by the coupling to an environment, its unitarity can be dropped, yielding more flexibility for implementing MIPTs with open systems. In particular, one might choose $\hat C^{(\pm)}_{l \,\text{odd}}=\hat h_{l\pm1} $, $\hat C_{l\, \text{even}} =\hat{\mathds{1}}$, i.e., $\hat L^{(\pm)}_{l\,\text{ odd}}= \hat c^\dagger_{l\pm1} \hat c_l$. In the continuum limit, these operators become $\pm (\partial_x \hat c^\dag_x) \hat c_x + \hat n_x$. The gradient terms lift the degeneracy of the dephasing operators $\hat n_l$ and, together with the Hamiltonian $\hat H$, provide a unique dark state. However, they do not modify universal properties due to their RG irrelevance.

\emph{MIPT and pre-selection in Rydberg arrays.} -- As a second realization of a pre-selected MIPT, we discuss an array of Rydberg atoms in the facilitation regime~\cite{Gutirrez_2017,Helmrich_2018,Festa_2022}. Each atom $l$ hosts a qubit described by Pauli operators $\hat X_l, \hat Y_l, \hat Z_l$. The dynamics of $L$ qubits is generated by the Hamiltonian~\cite{Marcuzzi2016,PhysRevLett.111.215305}
\begin{align}\label{eq:RydbergHam}
    \hat H=\sum_l \hat n_l(\hat X_{l-1}+\hat X_{l+1}), \text{ with } \hat n_l\equiv \tfrac12(\hat{\mathds{1}}+\hat Z_l).
\end{align}
At each time step $dt$, it acts like a controlled-$\exp(-idt \hat X)$ gate on neighboring qubits. This yields the dark state $|D\rangle=|00...00\rangle$, while any other state in the computational basis quickly approaches a volume law entangled state. A MIPT is ensured when the entanglement growth caused by the Hamiltonian is put into competition with disentangling measurements of the qubit orientation $\hat Z_l$, with scaling exponents consistent with two-dimensional bond percolation~\cite{PhysRevB.101.104301,Zabalo2020}, see Fig.~\ref{Fig3}.

In fact, a pre-selection quantum simulation scheme for this model has already been implemented in experiment (yet violating criterion (c3))~\cite{Gutirrez_2017,Helmrich_2018,Festa_2022}: Consider projective measurements of $\hat Z_l$, as shown in Fig.~\ref{Fig3}. The Hamiltonian possesses no particular symmetry, thus a natural choice to implement pre-selection is
\begin{align}
    \hat C_l=\hat X_l\to \hat L_l=|0\rangle\langle 1|_l\equiv \hat\sigma^-_l.
\end{align} 
While this operators steers towards $|D\rangle=|00...00\rangle$ as unique dark state, the  local operator $\hat X_l$ represents a relevant perturbation and is expected to change the universal behavior. Physically, $\hat L_l$ describes spontaneous emission. Indeed, facilitated Rydberg arrays subject to decay are suspected to undergo an absorbing state phase transition in the directed percolation universality class~\cite{Marcuzzi2016,Gutirrez_2017}. In order to make the correction RG irrelevant, meeting (c3), we propose the conditioned version
\begin{align} \hat C_{l,\pm}= (\hat{\mathds{1}}-\hat n_{l\pm1})\hat X_l\to \hat L^\pm_l =\hat\sigma_l^-(\hat n_l-\hat n_{l\pm 1})=\pm(\nabla\hat\sigma_l^-)\hat n_l.
\end{align}
While $\hat C_l$ is not unitary, implementation of $ \hat L_l $ in controlled quantum platforms~\cite{Weimer2010,Barreiro_2011,Chertkov_2022} seems in reach for the quantum simulation of a MIPT. 

\emph{Conclusion. } -- Pre-selection reveals the physical meaning of MIPTs in terms of quantum absorbing state phase transitions in representative wave functions. This result is rationalized by the individual measurement trajectories hosting the same information as the entire ensemble in a large system. In this sense, it parallels post-selection, but it avoids exponential overhead. On the practical side, this opens up a clear path towards experimental observability of MIPTs in controlled quantum platforms. On the theory side, it sparks characterizing novel non-equilibrium universality classes in the quantum regime. While we demonstrated the tight connection of stationary behaviors of MITPs and pre-selected scenarios, a particular challenge is posed by exploring their mutual relation in terms of dynamics.

\emph{Note added}: During completion of this manuscript we became aware of related work, discussing an example of a MIPT made observable by feedback~\cite{Wilson2021}.

\begin{acknowledgments}
  We thank E. Chertkov, A.~J.~Daley, M. Foss-Feig and J.~H.~Wilson for fruitful discussions. We acknowledge support from the Deutsche Forschungsgemeinschaft (DFG, German Research Foundation) under Germany's Excellence Strategy Cluster of Excellence Matter and Light for Quantum Computing (ML4Q) EXC 2004/1 390534769, and by the DFG Collaborative Research Center (CRC) 183 Project No. 277101999 - project B02. Furthermore we acknowledge support by the European Research Council (ERC) under the Horizon 2020 research and innovation program, Grant Agreement No. 647434 (DOQS). M.B. acknowledges funding via grant DI 1745/2-1 under DFG SPP 1929 GiRyd. The code for our numerical computations was implemented in Julia~\cite{bezanson17}. 
\end{acknowledgments}

\appendix
\section{Classical pre-selection for continuously monitored fermions}\label{app:continuous_monitoring}
Here, we provide the theoretical background for pre-selected monitored fermion dynamics in  the classical feedback variant, cf. Fig.~\ref{Fig2}. In order to continuously monitor a set of observables $\{\hat L_l^{(0)}\}$ with a rate $\gamma$, they are weakly coupled to an auxiliary system~\cite{Schomerus2019,Milburn1993,Yang2018}, which is then frequently measured~\cite{Steck2006}, i.e., on time scales much faster than the typical evolution time of the system. This reveals the so-called currents $ J_{l,t}=\langle \hat L_l^{(0)}\rangle_t+\tfrac{dW_l}{2\gamma dt}$, where $\langle \hat L_l^{(0)}\rangle_t=\Tr (\hat L_l^{(0)}\hat\rho_t^c)$ is the instantaneous expectation value, and $dt$ the typical time between successive measurements. Here $\hat \rho_t^c=|\psi\rangle\langle\psi|$ is the so-called conditioned state, i.e., the wave function conditioned on all previous measurement outcomes.
The randomness of the measurement outcome is encoded in the Wiener increment $dW_l$ with variance $\gamma dt$. In the absence of environmental dissipation, the quantum system remains in a pure state wave function $\ket{\psi}$, which is \emph{conditioned} on the stream of currents $\{J_{l,t}\}$. The temporal update of the conditioned state is  
\begin{align}\label{eq:DetMaster}    \partial_t\hat \rho_t^c=&i[\hat\rho_t^c,\hat H]-\tfrac{\gamma}{2}\sum_l[\hat L_l^{(0)},[\hat L_l^{(0)},\hat\rho_t^c]] \\   &+2\gamma\sum_l(J_{l,t}-\langle\hat L_l^{(0)}\rangle_t) \{\hat L_l^{(0)}-\langle \hat L_l^{(0)}\rangle_t,\hat\rho_t^c\}.\nonumber\end{align}
When the currents at each time step are known from the measurement read-out, the state update is \emph{deterministic}. It can, e.g., be simulated in real time parallel to the experiment for systems of low complexity, such as small systems or systems described by Gaussian wave functions. The latter thus provides a case when the instantaneous expectation values $\langle \hat{L}^{(0)}_l \rangle$ can be extracted in every time step from simulation. We use this insight in the classical pre-selection scheme for monitored free fermions that is discussed in the main text and Fig.~\ref{Fig2}, considering a state-dependent Hamiltonian $\hat{H} = \hat H [\hat \rho^c]=\hat H[\lbrace \langle \hat{L}^{(0)}_l \rangle \rbrace]$. 

An alternative scheme, e.g., applicable for more complex wave functions that cannot be simulated in real-time, may estimate the instantaneous expectation values directly from the measurement outcomes, i.e., the currents $J_{l,t}$, by applying appropriate filtering, such as using a Kalman~\cite{PhysRevLett.114.223601} or a low pass filter~\cite{Young_2021}. Advanced filtering schemes thus can represent an alternative way to resolve MIPTs in the classical feedback scheme. A further alternative is represented by  the quantum feedback and quantum simulation variants discussed in the main text, which do not require the classical inputs.

In order to simulate the free fermion setup as described in the main text in practice, we parameterize the conditioned state as $\hat{\rho}^c=\ket{\psi} \bra{\psi}$ with $\ket{\psi}_t = \prod_{l=1}^{L/2} \left( \sum_{j=1}^L U_{j,l}(t) \hat{c}_j^\dagger \right) \ket{0}$ (cf.~\cite{Cao2019,alberton2021enttrans}), and the $L \times L/2$ matrix $U$ contains normalized single-fermion wave functions. Up to leading order $\mathcal{O}(dt)$, the evolution of $U$ is computed by matrix multiplication
\begin{align}
    U(t+dt) =& \text{diag}(e^{dW_{1,t}+ \gamma dt (2 \langle \hat{n}_1 \rangle_t-1)}, \dots, e^{dW_{L,t}+ \gamma dt (2 \langle \hat{n}_L \rangle_t-1)}) \\ & \times e^{-i h[U(t)] dt} U(t). \notag
\end{align}
Here, we indicate that with pre-selection, the Hamiltonian matrix $h[U(t)]$ is itself a function of the state, i.e., the time-dependent matrix $U(t)$, with
$\langle \hat{n}_l \rangle_t = [U(t)U(t)^\dagger]_{l,l}$. It is a hopping matrix with entries $h_{l,l+1}[U(t)]=h_{l+1,l}[U(t)]= 2-(\lfloor\langle \hat n_{l-1}
\rangle\rfloor-\lfloor\langle\hat n_{l}\rangle\rfloor)^2-(\lfloor\langle \hat n_{l+1}
\rangle\rfloor-\lfloor\langle\hat n_{l+2}\rangle\rfloor)^2 =r_l[\hat{\rho}_t^c] $ and zero otherwise. This implements the classical pre-selection routine described in the main text. After each time step, for which we choose $dt=0.05$, normalization is implemented by performing a QR-decomposition $U=QR$ and setting $U=Q$. 

\begin{figure*}[t]
  \includegraphics[width=\linewidth]{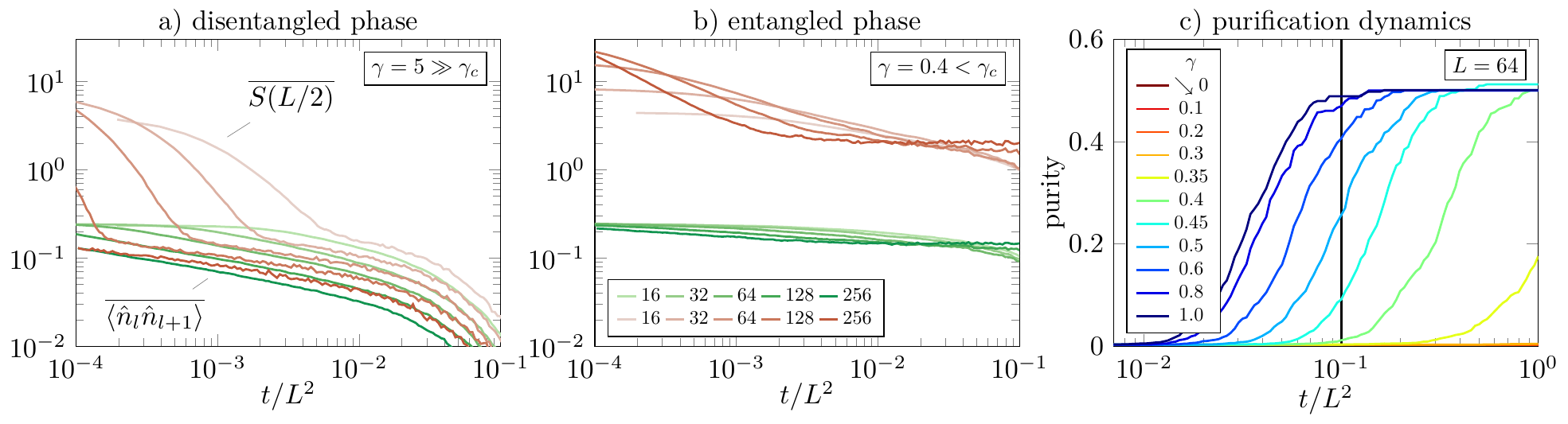}
  \caption{(a), (b) System-size dependence of the approach to the target state. For strong measurements $\gamma \gg \gamma_c$, the feedback pushes the system into the target state starting at $t \approx L^2/100$, independently of the precise value of $\gamma$ and for a wide range of $L$. At the same timescale, weak measurements $\gamma < \gamma_c$ lead to a plateau for the proximity to the target state and the entanglement entropy. For sufficiently large system sizes $L$, we observe no decay plateau of the plateau on algebraic time scales, which confirms our expectation of an exponentially long lifetime $t\sim\exp(L)$ of the active state. This means that we observe stationary behavior in both regimes at $t \approx L^2/10$ and we pick this timescale to study the stationary state properties in both regimes. For $\gamma \approx \gamma_c$, large $L$ are needed to clearly identify one of the two cases, which we treat by finite size scaling analysis at $t=L^2/10$. Before the stationary state is established, we observe a characteristic algebraic scaling regime, $\overline{\langle \hat{n}_l \hat{{n}}_{l+1} \rangle} \sim t^{-\sigma}$ with $\sigma \approx 1/3$ in the strongly measured regime. (c) For strong measurements the trajectory ensemble purifies at the same timescale, while weak measurements leave it fully mixed for exponentially long times. The final purity is $1/2$ due to the degeneracy of the target state.}
  \label{FigApp1}
\end{figure*}

Each trajectory is initialized in a random, half filled product state (e.g. $\ket{01101001\dots}$), which is evolved under an unconditioned Hamiltonian ($r_l=1$) for a short period ($t=-5$ to $t=0$) to reach an entangled intial wave function at $t=0$. Then, the classical pre-selection scheme is applied as described above. During the evolution, we compute the entanglement entropy $S(L/2)=-\Tr \hat{\rho}^c_{L/2} \log \hat{\rho}^c_{L/2}$ for the half-system reduced density matrix $\hat{\rho}^c_{L/2}=\Tr_{L/2} \hat{\rho}^c$, the order parameter $\langle \hat{n}_l \hat{n}_{l+1} \rangle=\Tr \hat{n}_l \hat{n}_{l+1} \hat{\rho}^c \geq 0$ and the ensemble purity (see below). 
Each observable reaches a (quasi-) stationary plateau at times $t \sim \mathcal{O}(L^2)$, as we show in Fig.~\ref{FigApp1}. The algebraic scaling with system size results from the diffusive motion of particles in number-conserving systems subject to decoherence. It has to be distinguished from the exponentially long, asymptotic time-scale $t\sim \exp(L)$, growing with the size of the Hilbert space, on which the system always falls into the absorbing state, see Fig.~\ref{FigApp1}. Throughout this work, we set the stationary values to be reached at $t=0.1L^2$, which turns out to yield consistent and converged expectation values, shown in Fig.~\ref{Fig2}. The observables at this timescale are extracted by averaging over many different trajectories (denoted by an overbar). Each result is then also time-averaged over the period $t=\frac{5}{6} L^2/10$ to $t=\frac{6}{5} L^2/10$ and fitted by linear regression, yielding the error estimates shown in Fig.~\ref{Fig2}.  

\section{Quantum pre-selection for projective measurements and quantum jumps}\label{app:quantum_preselection}
Here, we derive the quantum master equation for pre-selection, Eq.~\eqref{eq:modifiedmaster} in the main text. We consider two cases, projective measurements and a quantum jump evolution, which yield analogous results. The master equation is conveniently derived from a continuous evolution equation of the wave function, the stochastic Schrödinger equation (SSE)~\cite{Steck2006,alberton2021enttrans}. For each trajectory, we make use of the conditioned state $\hat\rho^c=|\psi\rangle\langle\psi|$, i.e., the pure state wave function conditioned on all measurement outcomes. For a set of local measurements $\hat L_l^{(0)}$ denoted by an index $l$, e.g., spin orientation $\hat L_l^{(0)}=\hat Z_l$ or fermion density $\hat L_l^{(0)}=\hat n_l$, we introduce the projector $\hat P_l$ on one measurement outcome, e.g., $\hat P_l=(\hat{\mathds{1}}+\hat Z_l)/2$ for spins and $\hat P_l=\hat n_l$ for fermions. We refer to the orthogonal projection as $\hat Q_l=\hat{\mathds{1}}-\hat P_l$. Without loss of generality, we assume that $\hat P_l$ is compatible with the dark state, such that $\hat P_l\hat\rho_D=\hat\rho_D$ and $\hat Q_l\hat\rho_D=0$.   At each time step $dt$, the conditioned state evolves according to $\rho^c_{t+dt}=\rho^c_t+d\rho^c_t$ with
\begin{align}\label{eq:condMasterProj}   d\hat\rho^c_t=&idt [\hat\rho^c_t,\hat H]+\sum_l dW_l\left(p_l\frac{\hat P_{l}\hat\rho^c_t \hat P_{l}}{\langle \hat P_{l}\rangle_t}+(1-p_l)\frac{\hat Q_{l}\hat\rho^c_t \hat Q_{l}}{\langle \hat Q_{l}\rangle_t}-\hat\rho^c_t\right).\end{align}
Whether or not a measurement at coordinate $l$ is performed at this time step is determined by the discrete, random increment $dW_l=0,1$. It has expectation value $\mathds{E}(dW)=\gamma dt$, reflecting an average measurement rate $\gamma$. Once a measurement is performed, the Born probabilities are implemented by the random number 
$p_l=0,1$. It selects the measurement outcome associated with $\hat P_l$ with probability $\mathds{E}(p_l)=\langle\hat P_l\rangle$. The statistical average of Eq.~\eqref{eq:condMasterProj} yields the master equation 
\begin{align}\label{eq:ProjMEQ}
    \partial_t\hat\rho=i[\hat\rho,\hat H]-\gamma\sum_l [\hat P_l,[\hat P_l,\hat\rho]]
\end{align}
for the wave function ensemble $\hat \rho$. Due to the double commutator, the projector $\hat P_l$ can be replaced by the observable $\hat L_l^{(0)}$ and an appropriate numerical prefactor, e.g. $\hat P_l\to \hat n_l$ for fermions and $\hat P_l\to \tfrac12\hat Z_l$ for spins.

With quantum pre-selection, we immediately correct each `wrong' measurement result with a unitary $\hat C_l\hat C_l^\dagger=\hat{\mathds{1}}$. This replaces $\hat Q_{l}\hat\rho^c \hat Q_{l}\to \hat C_l\hat Q_{l}\hat\rho^c \hat Q_{l}\hat C_l^\dagger$ in Eq.~\eqref{eq:condMasterProj} as outlined in the main text. This correction is state-independent and averaging Eq.~\eqref{eq:condMasterProj} with correction yields
\begin{align}
    \partial_t\hat\rho=i[\hat\rho,\hat H]-\frac{\gamma}{2}\sum_l\left( [\hat P_l,[\hat P_l,\hat\rho]]+\{\hat L_l^\dagger \hat L_l,\hat\rho\}-2\hat L_l\hat\rho\hat L_l^\dagger\right).\nonumber
\end{align}
Here, the modified Lindblad operators $\hat L_l=\hat C_l\hat Q_l$ are non-Hermitian. They implement the IF-THEN condition: if the state $\hat\rho$ is locally in the target dark state, then $\hat Q_l\hat \rho=0$ and consequently $\hat L_l\hat\rho=0$. If, however, $\hat Q_l\hat\rho\neq0$, then $\hat C_l$ rotates the state locally towards the dark state. We remark that the correction acts on each `wrong' measurement outcome and therefore in the master equation above, the rate $\gamma$ at which the Lindblad operators act is locked onto the measurement rate. This is in principle different for the quantum simulation scheme, where an individual rate for both processes might be set by the bath degrees of freedom.

An analogous scheme can be implemented for a quantum jump evolution protocol~\cite{Daley}. Here, each degree of freedom $l$ is coupled to a detector, which measures only one possible outcome of the observable. An example is the detection of a particle by light scattering: once a scattered photon is observed (the detector `clicks'), it indicates the presence of a particle. However, not observing a scattered photon does not rule out the presence of the particle. The pre-selection scheme works by performing an immediate correction after a wrong measurement outcome has been recorded, which in the quantum jump evolution can only be a `click'. In order to make contact to projective measurements above, we associat a `click' with the projector $\hat Q_l$. When the detector `clicks', which happens with a probability $\gamma dt \langle Q_l\rangle$, the state is projected by $\hat Q_l$. In the quantum jump protocol, the conditioned state follows the SSE \begin{align}d\hat\rho^c=&idt [\hat\rho^c,\hat H]+\sum_l \left[d\tilde W_l\left(\frac{\hat Q_{l}\hat\rho^c \hat Q_{l}}{\langle \hat Q_{l}\rangle}-\hat\rho^c\right)-\frac{\gamma dt}{2}\{\hat Q_l-\langle\hat Q_l\rangle,\hat\rho^c\}\right].\nonumber\end{align} Whether or not a `click' is recorded is determined by the discrete stochastic increment $d\tilde W_l=0,1$, which clicks with probability $\mathds{E}(d\tilde W_l)=\gamma dt\langle \hat Q_l\rangle$. Performing the statistical average over all measurement outcomes without pre-selection yields the master equation~\eqref{eq:ProjMEQ} with $\gamma\to\gamma/2$.

Pre-selection in the quantum jump evolution is implemented by modifying the projection of the state $\hat Q_{l}\hat\rho^c \hat Q_{l}\to \hat C_l\hat Q_{l}\hat\rho^c \hat Q_{l}\hat C_l^\dagger$ whenever a `click' has been recorded. This yields the master equation for the ensemble average
\begin{align}
    \partial_t\hat\rho=i[\hat\rho,\hat H]-\frac{\gamma}{2}\sum_l\left[\{\hat L_l^\dagger \hat L_l^{\phantom{\dagger}},\hat\rho\}-2\hat L_l^{\phantom{\dagger}}\hat\rho\hat L_l^\dagger\right].
\end{align}
It has the same rate and Lindblad operators as the ensemble evolution under projective measurements, but it does not feature the additional dephasing $\sim [\hat P_l,[\hat P_l,\hat\rho]]$.

\section{Mean-field treatment of fermions with quantum pre-selection}\label{app:meanfield}
A more intuitive picture of the dynamics of fermions in the quantum feedback and the quantum simulation scheme is obtained when mapping the Néel state to a vacuum state $|01010..\rangle\to|00000..\rangle$ by performing an exact particle-hole transformation $\hat c_l\leftrightarrow\hat c_l^\dagger$ on every even site $l$. This maps the conditioned Hamiltonian in the main text to
\begin{equation}\label{eq:condFerm2} 
\hat H\to \hat H_{\text{ph}}=\sum_l \hat n_l(\hat h_{l+1}+\hat h_{l-2})
\end{equation}
 with $\hat h_l=(\hat c_{l}\hat c_{l+1}+\text{H.c.})$. The Hermitian Lindblad operators are unchanged under this transformation since double commutators with $\hat n_l $ or $1-\hat n_l$ are identical. The non-Hermitian Lindblad operators on odd sites transform to $\hat L_l=\hat n_l(1-\hat n_{l+1})+i\hat c_l\hat c_{l+1}$, yielding the master equation
 \begin{align}
     \partial_t\hat\rho&=i[\hat\rho,\hat H_{\text{ph}}]-\gamma\sum_{l\text{ even}}[\hat n_l,[\hat n_l,\hat\rho]]\nonumber\\
     &-\frac{\gamma}{2}\sum_{l\text{ odd}}[\hat n_l,[\hat n_l,\hat\rho]]+\{\hat n_l, \hat\rho\}-L_l\hat\rho\hat L_l^\dagger.
 \end{align}
The Lindblad operators $\hat n_l$ compete with the coherent generation of particle and hole pairs in the Hamiltonian. The operators $\hat L_l$ guide the state towards the vacuum, once the Hamiltonian is sufficiently suppressed. The dark state is a translational invariant product state. Thus all correlation functions factorize according to Wick's theorem. Close to the dark state, we can thus apply a mean-field theory, which captures the state-evolution in terms of the homogeneous density $n\equiv \langle \hat n_l\rangle$ and the pair-correlation $K\equiv i\langle c_l c_{l+1}-\text{H.c.}\rangle$. The mean-field theory becomes exact once the dark state is reached and represents a reasonable approximation in its vicinity. It yields
\begin{align}
    \partial_t n=n(4K-\gamma n), \ \partial_t K=4n(1-2n)-4\gamma K.
\end{align}
Inserting the stationary solution for the pair-correlation function $K=n(1-2n)/\gamma$ provides an evolution equation for the density alone 
\begin{align}
    \partial_t n=\frac{n^2}{\gamma}(4-8n-\gamma^2)
\end{align}
It shows that $n=0$, i.e., the absorbing state, is always a stationary solution of the time evolution. Furthermore, the absorbing state is the only stationary solution for $\gamma\ge 2$, making it a robust dark state in this regime. For $\gamma<2$, however, any non-zero initial density $n>0$ will lead to an algebraic growth of $n$ towards the value $n=\tfrac12-\tfrac{\gamma^2}{8}$. This indicates a robust active phase with $n>0$ in this parameter regime. In summary, the mean-field analysis predicts both a robust active and a robust absorbing phase, separated by a critical point at $\gamma=\gamma_c=2$.

\section{Field theory for fermions with pre-selection}\label{app:fieldtheory}
For $U(1)$-symmetric fermions in one dimension, mean-field theory provides only a very rough understanding of the critical dynamics at a quantum phase transition. An improved description is obtained from a long-wavelength theory, where the critical point is approached from the weak measurement phase. In this regime, the one-dimensional fermions are gapless and therefore described more accurately in terms of a hydrodynamic approach, namely via bosonization. Here, we work in the original basis (not particle-hole transformed). The fermion density and current in the spatial continuum $l\to x$ are described by boson field operators $[\partial_x\hat\theta_x,\hat\phi_{x'}]=i\delta(x-x')$. In order to develop the effective long-wavelength description of the absorbing state phase transition of monitored fermions with pre-selection, we discuss the modifications to the Hamiltonian and the Lindblad operators in the bosonization framework.

In the fermionic description, the Lindblad operators $\hat L^{(0)}_l\to \hat L_l$ are modified in such a way that (i) $\hat L_l$ is non-Hermitian, (ii) the modification does not commute with $\hat L^{(0)}$ and (iii) it is irrelevant in the RG sense. The unmodified Lindblad operators in the continuum translate to $\hat L^{(0)}_x=(\hat L^{(0)}_x)^\dagger\sim \partial_x\hat\phi_x+\cos\hat\phi_x$ in bosonization~\cite{buchhold2021effective}. Here we expand the densities $\hat L^{(0)}_x$ around their mean field value $\hat L^{(0)}_x-\langle \hat L_x^{(0)}\rangle$. Measuring the densities aims to localize the particles via the backscattering term $\sim\cos\hat\phi_x$, pushing the system into eigenstates of the boson field $\hat\phi_x$. However, since any eigenstate $\hat\phi_x\hat\rho_\phi=\phi_x\hat\rho_\phi$ of the operator $\hat\phi_x$ solves 
\begin{align}
    \{\hat L_x^{(0)} \hat L_x^{(0)},\hat\rho_\phi\}-2\hat L_x^{(0)}\hat\rho_\phi\hat L_x^{(0)}=0,
\end{align}
the master equation evolves towards a statistical mixture of all possible localized states, which erases the information on individual wave functions.

With pre-selection, the Lindblad operators are modified such that criteria (i)-(iii) above are fulfilled. The modification $i\hat c_{l+1}^\dagger\hat c_l$ to the Lindblad operator in Eq.~\eqref{eq:modifiedmaster} has the form of the microscopic density of right-moving fermions, i.e., a current, with $\hat c_{l+1}^\dagger\hat c_l\to(1-i)\partial_x\hat\theta_x$. We readily see that by adding this RG-irrelevant current $\hat L_x^{(0)}\to \hat L_x=(1+i)\partial_x\hat\theta+\hat L^{(0)}_x$, the desired properties are met. With this modification, the maximally mixed state ceases to be a general solution of the master equation:
\begin{align}\label{eq:LuttingerMod}
    \int_x\{ \hat L^\dagger_x\hat L_x,\hat{\mathds{1}}\}-2\hat L_x\hat{\mathds{1}}\hat L_x^\dagger\sim -4\int_x \sin\hat\phi_x.
\end{align}
This simple result provides already an accurate estimation of the absorbing phase transition: at weak measurements, the non-linear operators $\sim \cos\hat\phi_x, \sin\hat\phi_x$ are RG-irrelevant at large distances~\cite{buchhold2021effective}. Thus Eq.~\eqref{eq:LuttingerMod} effectively becomes zero and the maximally mixed state remains robust. For strong measurements, however, the non-linear operators become relevant and so does Eq.~\eqref{eq:LuttingerMod}. The only allowed stationary state is then one where density fluctuations are pinned $\sin\hat\phi_x\to0$ in the entire ensemble. 

Approaching the transition from the weakly measured phase, the dynamics is dominated by the gapless Hamiltonian and the noise induced by the unmodified Lindblad operators $\hat L_x^{(0)}$. The unmodified hopping Hamiltonian in bosonization is
\begin{align}
    \hat H^{(0)} =\int_x (\partial_x\hat\phi_x)^2+(\partial_x\hat\theta_x)^2.
\end{align}
The modification $\hat H^{(0)}\to \hat H$ amounts to multiplying each hopping term with $\hat h_l\to (1+(-1)^l(\hat n_{l-1}-\hat n_{l+2}))\hat h_l$, which to leading order $\hat n_l\to \rho_0-\tfrac1\pi$ is an RG-irrelevant correction, as required by (c3). The statistical fluctuations, i.e., the noise, at the transition is equally determined by gapless operators 
\begin{align}
  \int_x  [\hat L_x^{(0)},[\hat L_x^{(0)},\hat\rho]]=\int_x \tfrac{1}{\pi^2}[\partial_x\hat \phi_x,[\hat\partial_x\phi_x,\hat\rho]].
\end{align}
The dynamics close to the transition is transparently revealed by the Heisenberg-Langevin equation of the density field $\hat \phi_x$. In order to derive it, we map the full master equation with pre-selection to a non-equilibrium Keldysh path integral~\cite{buchhold2021effective}. The equation of motion for the operators $\hat\phi_x$ can then be obtained by first integrating out the fields associated to the operators $\hat\theta_x$, and then expanding the action in the so-called quantum fields~\cite{Kamenev2011}. This yields to leading order 
\begin{align}
    (\partial_t^2-\partial_x^2)\hat\phi_x+\gamma\cos\hat\phi_x+\underbrace{\gamma\sin\hat\phi_x(\partial_t\hat\phi_x+\sin\hat\phi_x)}_{\text{pre-selection}}=\hat\xi_x,
\end{align}
with the quantum white noise $\hat\xi_x$, which has zero mean and variance (in momentum space) $\langle\hat\xi_k\hat\xi_{-k}\rangle=\gamma k^2$. Without pre-selection, this equation describes a conventional Luttinger liquid subject to heating or dephasing in the  density $\sim \hat n_l$~\cite{PhysRevLett.109.045302}. This term yields fluctuations $\sim k^2$ for each momentum mode $k$ and pushes the system into a maximally mixed state. With pre-selection, however, the field $\hat\phi_x$ experiences cooling $\sim\partial_t\hat\phi_x$ once the non-linear operator $\sin\hat\phi_x$ becomes relevant. This term dominates over the noise and pushes the system into the absorbing state. 

Overall, this gives rise of a new type of absorbing state phase transition, where the order parameter $\hat\phi_x$ remains gapless both in the spectrum $\sim \hat H$ as well as in the noise $\sim \hat L_x^{(0)}$ in the entire active phase. The absorbing state phase transition corresponds to a generation of a mass scale in the spectrum according to the BKT mechanism, which steers the state into the noiseless configuration $\hat\phi_x=\text{const.}$, i.e. at momentum $k=0$, which is blind to the noise $\sim k^2$ in Fourier space.

It is instructive to compare the present version of an absorbing state phase transitions to more conventional, classical variants. A paradigmatic example in the latter context is directed percolation, which we consider here for definiteness. The overarching common feature is the existence of a dark or absorbing state of the full master equation. Also common is a suppressed noise level in the vicinity of the dark state, which vanishes in the dark state itself, creating its `dark' or `absorbing' property. This is responsible for the existence of genuine phase transitions in $1+1$ dimensions, which can have no a counterpart at finite temperature thermal equilibrium, where the noise level is flat and non-zero. However, differences are present in the incarnations of the suppressed noise level: In the case discussed above, there is a dark state at $k=0$, while in the vicinity, there is an additive noise level which scales $\sim k^2$ as indicated above. This scaling of the noise level equips the system with an effectively $1+1$ dimensional phase space, analogous to a quantum problem at $T=0$ \cite{Marino2016}, and thus enabling a quantum phase transition (for comparison, a noise level scaling $\sim k^0$ would reduce the problem to an effectively $1+0$ dimensional,  classical one). In contrast, in classical absorbing state transitions, the dark state is realized by a noise  which is multiplicative but scales $\sim k^0$, i.e., the noise level scales with the order parameter field of the absorbing state transition. Therefore, it vanishes identically in the inactive phase, while being proportional to the order parameter in the active phase.

\section{Purity}\label{app:purity}
An convenient and intuitive measure for an absorbing state phase transition in a quantum system is the purity of the wave function ensemble $\tau \equiv\text{tr}\hat\rho^2$. For a pure state, $\hat\rho^2=\hat\rho$ and the purity $\tau=1$. Generally, the purity is proportional to the number of quantum states that contribute to the stationary ensemble, $\tau\sim |\mathcal{H}_{ss}|^{-1}$, where $\mathcal{H}_{ss}$ is the Hilbert space of stationary wave functions. 

Without pre-selection, the stationary state is maximally mixed $\hat\rho\sim\hat{\mathds{1}}$ and thus $\tau\sim \exp(-L)$ decays exponentially with the system size. When pre-selection is applied, and when the system reaches an absorbing state,  $\tau=1/N_D$, where $N_D$ is the number of possible dark states.
When $\hat\rho$ is a statistical mixture of $N$ independently measured wave functions $|\psi^{(l)}\rangle$, i.e., $\hat\rho=\frac{1}{N}\sum_{l=1}^N|\psi^{(l)}\rangle\langle\psi^{(l)}|$, then
\begin{align}
    \tau=\frac{1}{N^2}\sum_{l,l'=1}^N|\langle\psi^{(l)}|\psi^{(l')}\rangle|^2.
\end{align}

In Figure~\ref{FigApp1} we show the purity of monitored free fermions undergoing a MIPT with pre-selection for a system of $L=64$ sites. At the critical point, it jumps from $\tau\sim0$ in the weak monitored phase to $\tau=\tfrac12$ in the strongly measured, absorbing phase. This is precisely the degeneracy of the (two possible) Néel states. 
\bibliography{Ent}

\end{document}